\documentclass{article}
\usepackage[preprint]{spconf}
\usepackage{amsmath,graphicx}
\usepackage{soul,color}

\copyrightnotice{\copyright\ IEEE 2022}
\toappear{To appear in {\it Proc.\ SLT 2022, Jan 09-12, 2023, Doha, Qatar}}

\usepackage{graphicx}
\usepackage{multicol}

\usepackage{hyperref}
\hypersetup{
    colorlinks=true,
    citecolor=black,
}

\title{DAMAGE CONTROL DURING DOMAIN ADAPTATION FOR TRANSDUCER BASED AUTOMATIC SPEECH RECOGNITION}
\name{Somshubra Majumdar\textsuperscript{*}, Shantanu Acharya\textsuperscript{*}, Vitaly Lavrukhin, Boris Ginsburg}
\address{\{smajumdar, shantanua, vlavrukhin, bginsburg\}@nvidia.com}

\def\mycopyrightnotice{%
{\footnotesize 978-1-6654-7189-3/22/\$31.00~\copyright~2023 IEEE\hfill} 
\gdef\mycopyrightnotice{}
}

\newcommand\blfootnote[1]{%
  \begingroup
  \renewcommand\thefootnote{}\footnote{#1}%
  \addtocounter{footnote}{-1}%
  \endgroup
}

\begin{document}
%
\maketitle

\blfootnote{\mycopyrightnotice}
\def\thefootnote{*}\footnotetext{~Equal contribution.}

\vspace{-24pt}
\begin{abstract}

Automatic speech recognition models are often adapted to improve their accuracy in a new domain. A potential drawback of model adaptation to new domains is catastrophic forgetting, where the Word Error Rate on the original domain is significantly degraded. 
This paper addresses the situation when we want to simultaneously adapt automatic speech recognition models to a new domain and limit the degradation of accuracy on the original domain without access to the original training dataset. We propose several techniques such as a limited training strategy and regularized adapter modules for the Transducer encoder, prediction, and joiner network. We apply these methods to the Google Speech Commands and to the UK and Ireland English Dialect speech data set and obtain strong results on the new target domain while limiting the degradation on the original domain.

\end{abstract}

\section{Introduction}
\label{sec:introduction}

Using a pre-trained Automatic Speech Recognition (ASR) system on a different domain than the one it was trained on, usually leads to severe degradation in Word Error Rate (WER). The adaptation of end-to-end ASR models to new domains presents several challenges. First, obtaining large amounts of labeled data on a new domain is expensive. Secondly, the most common domain adaptation approach is to fine-tune the ASR model, however, fine-tuning the model on relatively small amounts of data causes it to overfit the new domain. Finally, during adaptation, the WER of the model on the original domain may deteriorate, a phenomenon known as Catastrophic Forgetting  \cite{mccloskey1989catastrophic}. This is a significant drawback as the adapted model can no longer accurately transcribe speech from the original domain.

Prior works addressing domain adaptation generally fall into two categories: post-training adaptation and on-the-fly adaptation \cite{Sathyendra2022contextualadapter}. Post-training adaptation generally involves using domain-specific Language Models (LMs) \cite{Dingliwal2021DomainPT}. These models do not require the acoustic model to be re-trained but their usefulness is limited only to those applications where new domains differ only by new vocabulary terms that were not present originally. If we move to applications where the new domain differs by the speaker accent or new grammar, then these approaches do not perform well \cite{Turan2020AchievingMA}. On-the-fly adaptation techniques usually involve either Continual Joint Training (CJT) \cite{eeckt2021continual} or finetuning an existing pre-trained model. Since both these approaches require the training of the entire original model, they tend to require significant compute resources and data to perform well. Continual joint training has several drawbacks, primarily that it assumes that the entire original dataset is available for adaptation, and it does not consider the cumulative cost of training on an ever-growing dataset. 

\textit{Zhao et. al.} \cite{Zhao2021AUS} propose a unified speaker adaptation approach that incorporates a speaker-aware persistent memory model and a gradual pruning method. \textit{Hwang et. al.} \cite{hwang2022large} utilize the combination of self- and semi-supervised learning methods to solve unseen domain adaptation problems in a large-scale production setting for an online ASR model. While these approaches help in overcoming catastrophic forgetting, they make an implicit assumption that the data of the original domain is always available during the adaptation process. This assumption might not viable in practical scenarios, as production systems are generally trained on licensed or sensitive data which makes data sharing unfeasible.

\textit{Houlsby et. al.} \cite{houlsby2019parameter} propose adapter modules that are small sub-networks injected into the layers of a pre-trained neural network. The parameters of the pre-trained network are frozen and only the injected parameters are updated on the new domain. Even with just a small fraction of the entire model being trained on the new domain, adapters show performance comparable to fine-tuning. \textit{Tomanek et. al.} \cite{tomanek2021residual} demonstrate ASR domain adaptation by attaching adapters to the encoder part of a Recurrent Neural Network Transducer (RNN-T) \cite{graves2012transducer} and the Transformer Transducer (T-T) \cite{zhang2020transformer}. 

\textit{Eeckt et al.} \cite{eeckt2022cfadapters} use task-specific adapters to overcome catastrophic forgetting in domain adaptation. They demonstrate three adaptation techniques: (1) keeping the original model's parameters frozen (2) training with special regularization such as Elastic Weight Consolidation (EWC) \cite{kirkpatrick2017overcoming} (3) using Knowledge Distillation (KD) \cite{hinton2015distilling}. Still, the underlying assumption is that the original dataset is available during the adaptation.

We consider a \textit{Constrained Domain Adaptation} task, where the adaptation for the new domain is done without access to any of the original domain data.  We also strictly limit the allowed degradation on the original domain after adaptation \cite{eeckt2022cfadapters}.
Our main contributions are the following:
\begin{enumerate}
    \item We add adapter modules \cite{houlsby2019parameter} to the encoder, decoder, and joint modules of the  Conformer Transducer.  
    \item We train the adapters using various regularization techniques alongside a constrained training schedule, and show considerable improvement on the new domain while limiting degradation on the original domain.
    \item Finally, we propose a scoring scheme to select models that perform well in the constrained adaptation setting and evaluate the proposed approach on the Google Speech Commands \cite{warden2018speech} benchmark and the UK and Ireland English Dialect speech dataset \cite{demirsahin-etal-2020-open}.
\end{enumerate}

\section{Constrained Domain Adaptation}
\label{sec:adaptation}

\begin{figure*}[t!]
	\centering
	
	\begin{minipage}[b]{\linewidth}
		\centering
		\centerline{\includegraphics[width=\linewidth]{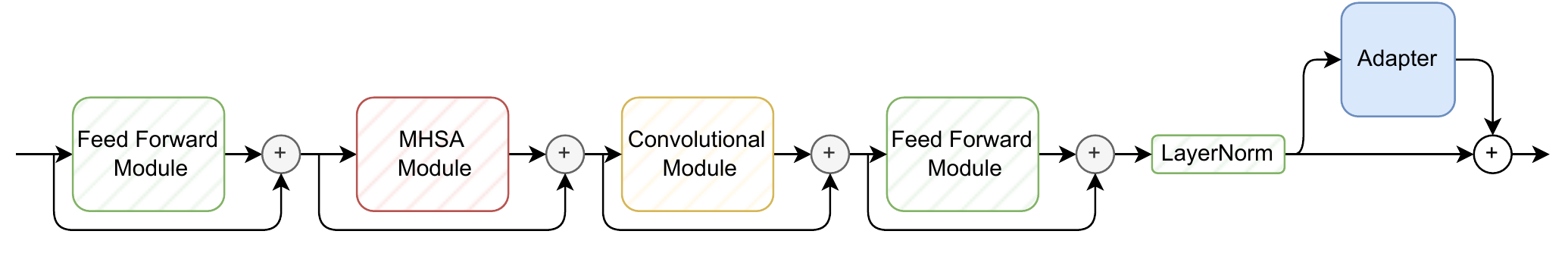}}
		\centerline{(a) Conformer Encoder Block}\medskip
	\end{minipage}
	\hfill
	\begin{minipage}[b]{0.3\linewidth}
		\centerline{\includegraphics[height=6.0cm]{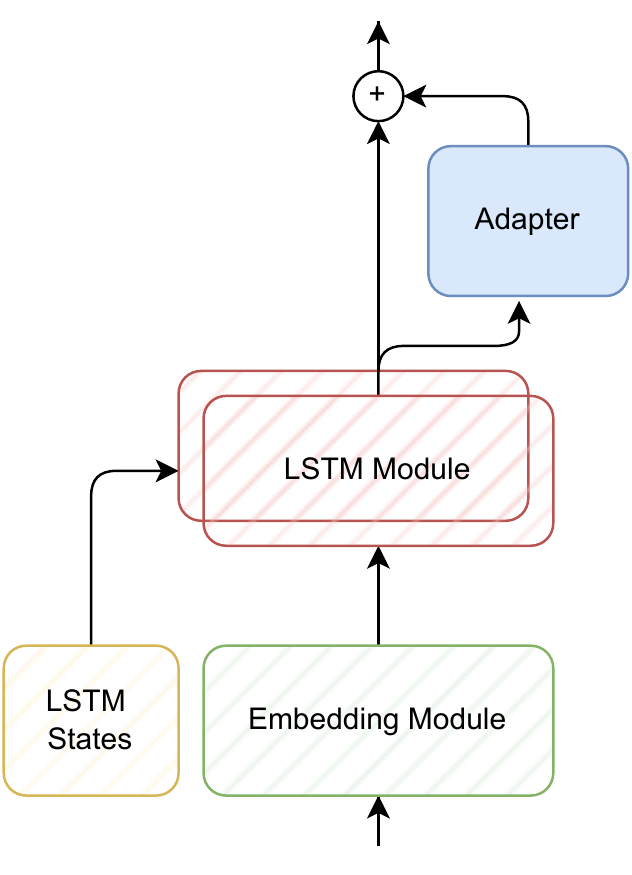}}
		\centerline{(b) Transducer Decoder Network}\medskip
	\end{minipage}
	\begin{minipage}[b]{0.3\linewidth}
		\centerline{\includegraphics[height=6.0cm]{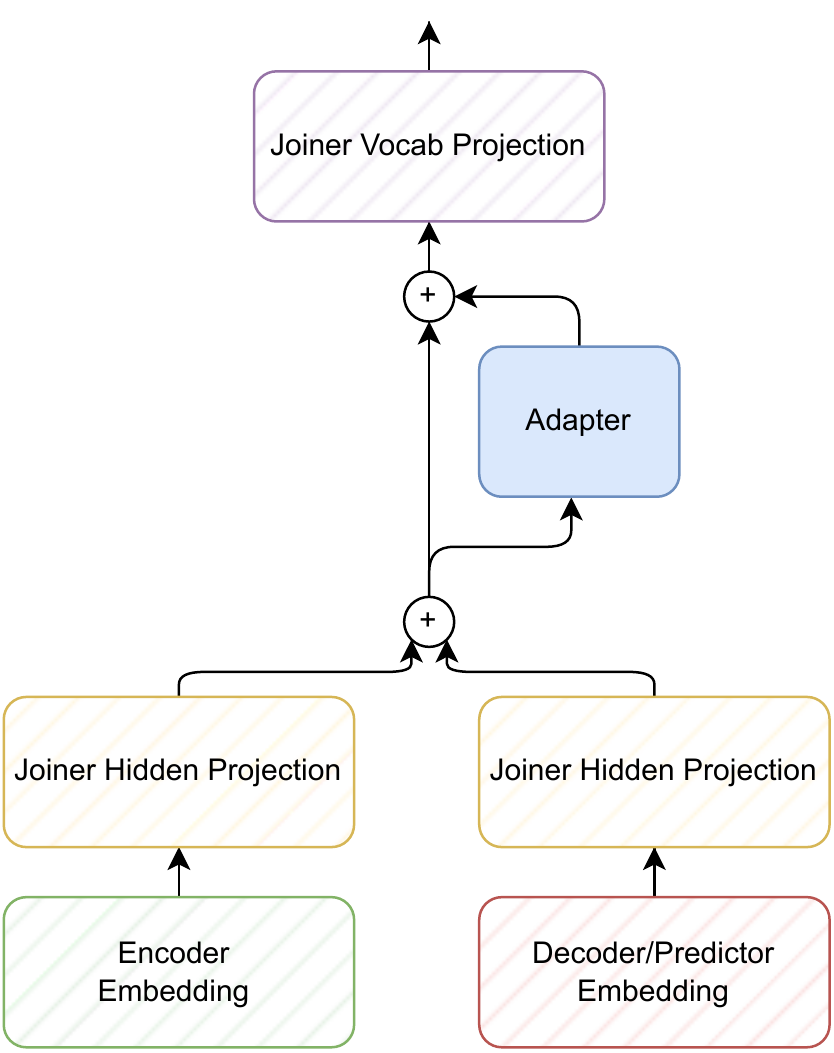}}
		\centerline{(c) Transducer Joint Network}\medskip
	\end{minipage}
	\begin{minipage}[b]{0.3\linewidth}
		\centerline{\includegraphics[height=6.0cm]{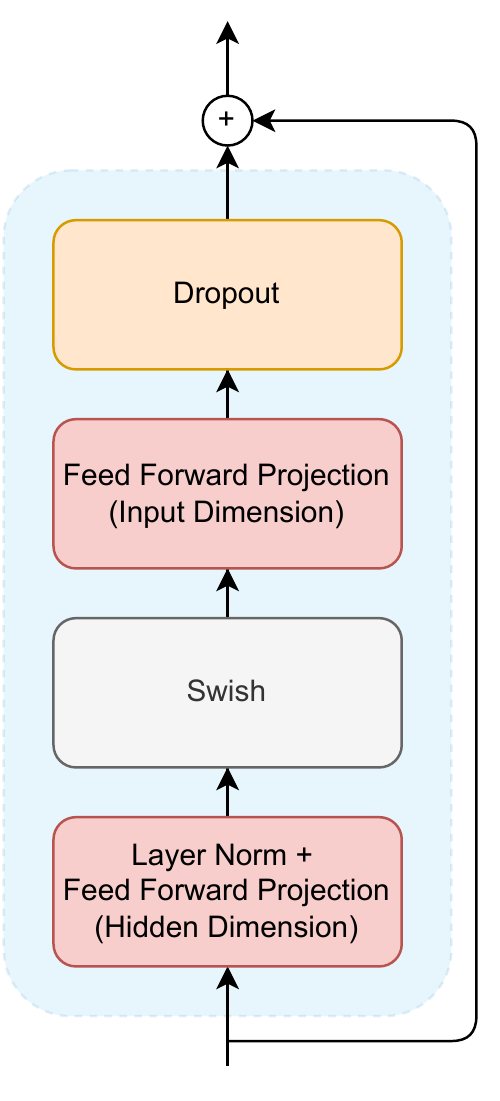}}
		\centerline{(d) Adapter Module}\medskip
	\end{minipage}
	\caption{Transducer with Adapters: (a) Conformer Encoder with Adapter . (b) Long Short-Term Memory Recurrent Neural Network-based Transducer Decoder network with Adapter . (c) Transducer Joint network with Adapter (d) Adapter module. Adapter modules comprise an initial Layer Normalization layer, followed by a linear projection to  hidden dimension, Swish activation \cite{ramachandran2018swish} and finally a linear projection to the input channel dimension. Note that due to stochastic depth regularization, the output of individual adapter modules can be skipped during training. Only the Adapter modules are updated during training, all other modules with hatch pattern are frozen.}
	\label{fig:positional_adapter}
\end{figure*}

\subsection{Degradation control on original domain}

In order to reduce the accuracy loss on the original domain, we formalize the first constraint as follows. During constrained domain adaptation, \textit{a candidate solution ($C$) must be evaluated on some evaluation set of the original domain prior to adaptation ($o$) and after adaptation ($o^*$) so as to limit the absolute degradation of WER to at most $\kappa$, where $\kappa$ is some predetermined acceptable degradation in WER on the evaluation datasets of the original domain}. 

To formalize the above constraint, we first define Word Error Rate (WER) degradation after adaptation as:
\begin{equation}
    \label{eqn:wer_degradation}
    \text{\textit{WERDeg}}_{o} = \max(0, \text{\textit{WER}}_{o^*} - \text{\textit{WER}}_o ) 
\end{equation}
where the subscript $o^*$, $a^*$ represent evaluation on the original domain and adapted domain after the adaptation process, and $o$, $a$ represent the evaluation on the original domain and adapted domain prior to the adaptation process respectively.

We then define the weight of degradation on the original domain as $O_{\text{SCALE}}$ , which is a scaling factor computed as :
\begin{equation}
    \label{eqn:original_scale}
    O_{\text{SCALE}}\ =\ \frac{1}{N}\sum _{i=1}^{N} \max \left( 0, \frac{\kappa_i - \text{\textit{WERDeg}}_{o,i}}{\kappa_i}  \right) 
\end{equation} 
where \textit{N} is the number of evaluation datasets from the original domain that the model was initially trained on and $\kappa$ is the maximum tolerable absolute degradation of word error rate on the original domain.

Next, we define relative WER improvement (WERR) on the new domain as $A_{\text{WERR}}$, such that 
\begin{equation}
    \label{eqn:adaptation_werr}
    A_{\text{WERR}}\ =\ \max \left( 0,\frac{\text{\textit{WER}}_{a} - \text{\textit{WER}}_{a^{*}}}{\text{\textit{WER}}_{a}}\right)
\end{equation} 
In this formulation, we only accept candidates which improve in WER on the new domain.

Combining the above definitions, we propose the following candidate selection metric which when maximized, yields candidates that maximize the relative WER improvement on the new domain, while simultaneously minimizing the degradation on the old domain. We define this metric as a score function in Eqn \ref{eqn:selection}: 
\begin{align}
    \label{eqn:selection}
    Score\ &=\ O_{\text{SCALE}}  *\ A_{\text{WERR}}  
\end{align}

We select the value of $\kappa$ to be 3\%, such that the absolute increase in WER on the original dataset is constrained to 3\%. It is possible to select a stricter threshold, however, the number of candidate solutions that satisfy the constraints decrease significantly, and exceedingly few valid candidates exist for the fine-tuning case. This score is maximized when the candidate attains the largest relative WER improvement on the new domain after scaling by a factor in the range of $[0,1]$ which indicates the weight of degradation of WER on the original domain. Note that the score has a minimum value of 0, if the absolute degradation of WER on the original domain surpasses $\kappa$, or if WER on the new domain becomes worse after adaptation. 

\subsection{Adaptation without access to original dataset}

During constrained domain adaptation, \textit{a candidate solution must only use the data of the new domain for adaptation, without access to any data from the original domain. It may use data from the original domain only for evaluation, in order to determine the severity of degradation on that domain after adaptation}.
When applying this constraint, we cannot freely compute the Coverage (COV) metric \cite{eeckt2022cfadapters} since it computes the difference between fine-tuning and CJT, though we may still utilize Learning Without Forgetting (LWF) \cite{li2017learning} which distills the model's knowledge using just the data of the new domain.

\section{Speech Transducers with adapters}
\label{sec:adapters}

Adapters are small sub-networks injected into specific layers of a pre-trained neural network, such that the original parameters of the network are frozen, avoiding gradient updates, and only the injected parameters of the adapter sub-networks are updated. 
Original adapters from \cite{houlsby2019parameter} are two-layer residual feed-forward networks, with an intermediate activation function, usually, ReLU \cite{nair2010relu}. Optionally adapters have a layer normalization applied to the input or output of the adapter.
Adaptors have been also applied to multilingual ASR \cite{winata2021adaptandadjust}, cross-lingual ASR \cite{wenxin2022crosslingualadapter}, self-supervised ASR \cite{bethan2022ssladapter} and contextual ASR \cite{Sathyendra2022contextualadapter}.  
Adapters have been found useful in reducing Catastrophic Forgetting in ASR \cite{eeckt2022cfadapters}.

Following an unified framework for adapters proposed by \textit{He et al.} \cite{he2022unified}, we consider three types of adapters for Transducer-based ASR models \cite{graves2012transducer}: 
(1) adapters added to the encoder of ASR networks (A-Enc), 
(2) the Decoder (Prediction Network) adapter (A-Dec),
(3) the Joint adapter (A-Joint). 
Figure \ref{fig:positional_adapter} details the configuration of adapters to the encoder, decoder, and joiner networks that comprise a standard Transducer ASR model, and details the construction of the adapter module itself.

The location of adapters should be determined by the adaptation task. In a transducer architecture, acoustic information can be adapted via encoder adapter, vocabulary adaptation can be done via decoder adapters, and joint acoustic and text adaptation can be done via the joint adapter. Note that decoder and joint adapters have less expressivity than their encoder counterpart, as they modify only a single layer rather than each encoder block. However, this limitation also tends to reduce the effect of catastrophic forgetting on the original domain. In order to enable a fair comparison between the adapters of different positions, we utilize larger hidden dimensions for the decoder and joint adapters, such that the total number of parameters added to the model is similar to the encoder adapter.

In general, adapter sub-networks add a small number of parameters to the original model, usually 1-2 \% of the total model parameters. Their size can be controlled by the hidden dimension of the feed-forward networks. Under the constrained domain adaptation scenario, we find that limiting the adapter size to roughly $0.25 - 0.5 \%$ of the original parameter count helps to prevent too rapid adaptation to the new domain, which impairs the model's accuracy on the original domain.

\begin{table*}[t]
    \centering
    \caption{WER ($\%$) on the UK and Ireland English Dialect speech dataset. $+$ indicates the result of unconstrained adaptation on the dataset, and $*$ indicates the result after constrained adaptation. The WER on the Librispeech Test Other set is averaged over all groups. \textbf{Bolded} cells denote the approach that obtains the maximum score computed via Eq. \ref{eqn:selection} for that group.}
    \vspace{4pt}
    \label{tab:slr83}
    \scalebox{0.925}{
    \begin{tabular}{ccccccccccccc}
        \hline
            \textbf{Model} &
            \textbf{Avg Test} &
            \textbf{Irish} &
            \textbf{Mid} &
            \textbf{Mid} &
            \textbf{North} &
            \textbf{North} &
            \textbf{Scott} &
            \textbf{Scott} &
            \textbf{South} &
            \textbf{South} &
            \textbf{Welsh} &
            \textbf{Welsh} \\
            &
            \textbf{Other} &
            \textbf{Male} &
            \textbf{Female} &
            \textbf{Male} &
            \textbf{Female} &
            \textbf{Male} &
            \textbf{Female} &
            \textbf{Male} &
            \textbf{Female} &
            \textbf{Male} &
            \textbf{Female} &
            \textbf{Male} \\
        \hline

        \textbf{Base} & 5.11 & 20.69 & 9.61 & 11.25 & 11.11 & 10.18 & 12.31 & 11.94 & 9.70 & 10.22 & 8.51 & 11.46 \\
        \hline
        \hline

        \textbf{FT}$^+$ & 7.68 $\pm$ 1.46 & 11.31 & 9.29 & 9.57 & 7.71 & 7.11 & 8.45 & 6.62 & 4.51 & 4.29 & 4.67 & 7.18 \\
        \textbf{A-Enc}$^+$ & 6.69 $\pm$ 0.35 & \textbf{13.79} & \textbf{8.56} & \textbf{9.16} & 9.19 & 9.07 & \textbf{9.83} & \textbf{8.02} & 8.24 & 7.88 & 7.02 & 9.42 \\
        \textbf{A-Dec}$^+$ & 5.51 $\pm$ 0.09 & 19.17 & 9.61 & 10.31 & 9.62 & 8.89 & 11.31 & 9.80 & 7.50 & \textbf{7.78} & 6.87 & 9.70 \\
        \textbf{A-Joint}$^+$ & 6.05 $\pm$ 0.18 & 17.79 & 9.05 & 9.64 & \textbf{9.19} & \textbf{8.29} & 11.33 & 9.39 & \textbf{7.03} & 7.36 & \textbf{6.34} & \textbf{9.25} \\
        \hline
        \hline

        \textbf{FT}$^*$ & 7.11 $\pm$ 0.48 & 11.58 & 8.64 & 9.50 & 8.34 & 7.19 & 8.85 & 6.62 & 4.55 & 4.65 & 4.79 & 6.93 \\
        \textbf{A-Enc}$^*$ & 5.65 $\pm$ 0.10 & \textbf{15.86} & \textbf{8.40} & \textbf{9.43} & 9.49 & 9.16 & \textbf{10.00} & \textbf{8.68} & 8.54 & 8.27 & 7.25 & \textbf{9.70} \\
        \textbf{A-Dec}$^*$ & 5.46 $\pm$ 0.09 & 19.52 & 9.53 & 9.77 & \textbf{9.33} & 8.94 & 11.26 & 9.92 & 7.71 & 7.91 & 6.78 & 9.89 \\
        \textbf{A-Joint}$^*$ & 5.40 $\pm$ 0.05 & 18.76 & 8.80 & 9.50 & 9.59 & \textbf{8.54} & 10.83 & 9.68 & \textbf{7.73} & \textbf{7.90} & \textbf{6.64} & 9.87 \\
        \hline
    \end{tabular}
    }
\end{table*}

\section{Experiments}
\label{sec:experiments}

The experiments were created using NVIDIA NeMo\footnote{\href{https://github.com/NVIDIA/NeMo}{\textcolor{blue}{https://github.com/NVIDIA/NeMo}}} \cite{kuchaiev2019nemo}. We fine-tune Conformer with location-specific adapters on two datasets: \textit{Google Speech Commands} dataset \cite{warden2018speech}  and on \textit{the UK and Ireland English Dialect speech dataset} \cite{demirsahin-etal-2020-open}. We select these datasets due to the significant distribution shift with respect to LibriSpeech. The Conformer-Transducer Large \cite{gulati2020conformer} is utilized as the base model, pre-trained on the LibriSpeech \cite{librispeech} corpora containing 960 hours of labeled English speech. The baseline is trained for 1000 epochs, at a global batch size of 2048 with the standard recipe described in \cite{gulati2020conformer}.

In an effort to prevent rapid deterioration of WER on the original domain, we incorporate dropout \cite{srivastava2014dropout} or stochastic depth \cite{huang2016deep} regularization to each individual adapter sub-network. We find that even though the number of parameters added is very small, dropout and stochastic depth play an important role in limiting the effect of catastrophic forgetting of the original domain. We use the AdamW optimizer \cite{loshchilov2018adamw} with 10 \% of training steps used for warmup, followed by the decay schedule proposed in  \cite{transformer}.

In order to select the hyperparameters for the adapter candidate solutions, a grid search was run over all possible combinations of adapter dimensions, dropout rate, stochastic depth, and training steps. Similarly, for the finetuning experiments, we compute a grid search over all combinations of the learning rate and the number of training steps. We consider an unconstrained candidate as one that maximizes the average WER on the new domain, without restriction on the WER obtained on the original domain after adaptation.

\subsection{Accent and Dialect Adaptation}
\label{sec:lang_transfer}

The UK and Ireland English Dialect speech dataset \cite{demirsahin-etal-2020-open} contains audio of English sentences recorded in 11 different dialects. The total audio duration for each group within the dataset ranges from 28 minutes to 7.5 hours. It is a particularly challenging dataset for the purpose of domain adaptation, as the minute amount of data per group allows any adaptation process to rapidly overfit to the new domain, significantly damaging accuracy in the original domain.

Due to high variability in the total duration of the available speech of all the dialect groups, the validation and the test set split for each group were decided by assigning them to 3 categories.  Groups with a total duration of less than an hour are assigned 5 minutes of audio for the validation set and 10 minutes of audio for the test set. Groups with a total duration of less than 3 hours are assigned validation and test sets with at most 15 minutes and 30 minutes of speech respectively. Finally, all groups with data surpassing 3 hours of speech are assigned 30 minutes for validation and 1 hour for test sets respectively.

For the finetuning and adapter experiments, we find learning rate of 1e-4 gave the best results. The model trained for 5000 steps obtains the largest score computed via Eq. \ref{eqn:selection}, which we use for the constrained adaptation task while training for 10000 steps had the lowest WER overall, which we select for the unconstrained adaptation task. Encoder adapters have a hidden dimension of 32 while the decoder and joint adapters have a hidden dimension size of 512. For the constrained adaptation task: (1) the encoder adapter was trained for 15000 steps with a stochastic depth chance of 90\%, (2) the decoder adapter was trained for 10000 steps with a stochastic depth chance of 50\%, and (3) the joint adapter was trained for 15000 steps with stochastic depth chance of 50\%. For unconstrained adaptation, all the adapters were trained on 15000 steps with: (1) encoder adapter having dropout with a chance of 90\%, (2) decoder adapter having stochastic depth chance of 50\%, and (3) the joint adapter having a stochastic depth chance of 20\%.

As can be seen from the result in Table \ref{tab:slr83}, even after we select candidates that limit degradation on the original domain by a maximum value of $\kappa$, unconstrained domain adaptation shows increased degradation on the original domain, significantly increasing the WER, while also attaining the best results on the new domain. It can also be seen that encoder adapters obtain similar results while utilizing only a fraction of the total number of parameters. We also note that in several cases, decoder and joint adapters are able to adapt to new dialects without substantial loss of WER on LibriSpeech. Decoder adapters primarily affect language modeling within a transducer which can provide a reasonable explanation for the limited improvement of such adapters for dialect transfer, where acoustic features are the source of the distribution shift. It can be seen that joint adapters often surpass decoder adapters since they are able to adapt to both acoustic and language features.

\vspace{-5pt}
\subsection{Speech Commands}
\label{sec:speech_commands}

The Speech Commands dataset \cite{warden2018speech} comprises one-second long utterances of 35 English words. The dataset is split into 3 parts: training, validation, and test containing 31 hours, 2.5 hours, and 3 hours of audio respectively. Although this dataset is generally used for speech classification tasks, it can be useful to make an ASR system support keyword recognition with higher accuracy.

For the finetuning and adapter experiments, we find a learning rate of 1e-5 with 1000 training steps had the largest score computed via Eq. \ref{eqn:selection}, making it the best candidate for the constrained adaptation task whereas a learning rate of 1e-4 with 600 training steps had the highest accuracy overall, which we select for unconstrained adaptation. Encoder adapters have a hidden dimension of 16 while the decoder and joint adapters have a hidden dimension size of 256. For constrained adaptation, all adapters were trained with a stochastic depth chance of 90\%. We train the encoder and decoder adapters for 600 steps and the joint adapters for 5000 steps in the constrained setting. For unconstrained adaptation, the hyperparameters for the decoder were the same as its constrained adaptation counterpart since the hyper-parameters having the highest accuracy also had the highest score computed via Eq. ~\ref{eqn:selection}. The encoder adapter was trained for 10000 steps with a stochastic depth chance of 90\% and the joint adapter was trained for 600 steps with a dropout of 90\%.

Table \ref{tab:gsc} shows accuracy after adaptation on the Speech Commands dataset. Since ASR models are trained on segments roughly 15-20 seconds long, global context models require substantial context in order to accurately transcribe speech. This dataset is challenging since it contains single-word utterances of at most 1 second in duration and therefore the ASR model must predict the correct label given insufficient context. This effect is also seen within adapters - even though all labels within the speech commands dataset are present in the original domain, training only the decoder adapter does not significantly improve the accuracy since the distribution shift is primarily acoustic in nature. The joint adapters are able to improve on the new domain to some extent since they jointly model acoustic and label sequence information unlike the decoder adapters, however, their improvements are minor as compared to encoder adapters that are primarily involved with acoustic modeling.

\begin{table}[thb]
    \centering
    \caption{Accuracy ($\%$) on the Speech Commands dataset. $+$ indicates the result of unconstrained adaptation on the dataset, and $*$ indicates the result after constrained adaptation. The WER on the LibriSpeech Test Other set is averaged over all subsets. \textit{Italicized} cells indicates violation of selection criteria defined by Eq. \ref{eqn:selection}, while \textbf{Bolded} cells denote the approach that obtains the maximum score computed for that group, when the selection criteria is not violated. All runs are averaged over 5 trials, with mean score selected for Speech Command subsets.}
    \vspace{4pt}
    \label{tab:gsc}
    \scalebox{0.925}{
    \begin{tabular}{ccccc} 
        \hline
            \textbf{Model} &
            \textbf{Test Other}($\downarrow$) &
            \textbf{10 Hr} ($\uparrow$) & 
            \textbf{20 Hr} ($\uparrow$) & 
            \textbf{30 Hr} ($\uparrow$) \\

            & 
            \textbf{WER} &
            \textbf{Acc} & 
            \textbf{Acc} & 
            \textbf{Acc} \\
        \hline

        \textbf{Base} & 5.12 & \multicolumn{3}{c}{$\underbrace{\qquad\qquad\quad \text{60.07} \qquad\qquad\quad}$} \\
        
        \hline
        \hline

        \textbf{FT}$^+$ & \textit{51.08} $\pm$ \textit{2.98} & 96.06 & 96.11 & 96.07 \\
        \textbf{A-Enc}$^+$ & \textit{18.17} $\pm$ \textit{0.43} & 93.36 & 93.35 & 93.59 \\
        \textbf{A-Dec}$^+$ & 6.75 $\pm$ 0.63 & \textbf{63.15} & \textbf{63.56} & \textbf{62.62} \\
        \textbf{A-Joint}$^+$ & \textit{40.16} $\pm$ \textit{1.38} & 81.28 & 80.91 & 80.93 \\
        \hline
        \hline
        
        \textbf{FT}$^*$ & \textit{29.78} $\pm$ \textit{0.57} & 93.20 & 93.30 & 93.30 \\
        \textbf{A-Enc}$^*$ & 6.24 $\pm$ 0.07 & \textbf{81.08} & \textbf{80.55} & \textbf{80.87} \\
        \textbf{A-Dec}$^*$ & 6.75 $\pm$ 0.63 & 63.15 & 63.56 & 62.62 \\
        \textbf{A-Joint}$^*$ & 5.79 $\pm$ 0.29 & 68.34 & 68.10 & 67.83 \\
        \hline
    \end{tabular}
    }
\end{table}

\section{Conclusion}
\label{sec:conclusion}

In this work, we discuss the adaptation of an end-to-end ASR model to a new domain without using any data from the original domain. We propose several techniques such as a limited training strategy and regularized Adapter modules for the Transducer encoder, prediction, and joiner network, which obtain strong results on a new domain without exhibiting significant degradation on the original domain. We provide a simple method to select models that satisfy the constraint on absolute degradation of WER on the original domain, while also maximizing the relative improvement on the new domain. When we apply these methods on the Google Speech Commands and UK and Ireland English Dialect speech data set, we observe strong results comparable to unconstrained fine-tuning. 

\section{Acknowledgement}
\label{sec:acknowledgement}

We thank Elena Rastorgueva, Jagadeesh Balam, Taejin Park and our colleagues at NVIDIA for feedback.

\bibliographystyle{IEEEbib}
\bibliography{references}

\end{document}